\documentclass[aps,pra,reprint,superscriptaddress]{revtex4-2}
\usepackage{amsmath,amssymb,graphicx,bm,braket,physics}
\usepackage[colorlinks=true,linkcolor=blue,citecolor=blue,urlcolor=magenta]{hyperref}
\usepackage{hyperref}
\usepackage{color}
\usepackage{mathtools}
\usepackage{bm}
\usepackage{enumitem}
\usepackage{graphicx}
\usepackage{epsfig}

\begin{document}

\title{Measurement Geometry as a Resource for Certifying Network Nonlocality}

\author{Leon Adachi}
\affiliation{Department of Astronomy and Geophysics, Faculty of Science, Tohoku University, Sendai 980-8578, Japan}
\affiliation{Frontier Research Institute for Interdisciplinary Sciences, Tohoku University, Sendai 980-8578, Japan}

\author{Le Bin Ho}
\affiliation{Frontier Research Institute for Interdisciplinary Sciences, Tohoku University, Sendai 980-8578, Japan}
\affiliation{Department of Applied Physics, Graduate School of Engineering, Tohoku University, Sendai 980-8579, Japan}

\date{\today}

\begin{abstract}
Quantum networks can exhibit nonclassical correlations that cannot be explained by classical models with independent sources. While the role of entanglement is well understood, the impact of measurement design remains largely unexplored. Here we develop an operational framework for certifying network nonlocality in the bilocal Alice--Bob--Charlie network using ancilla-assisted meters to evaluate the nonlocal observables required for bilocal and fully network nonlocal (FNN) witnesses. The approach successfully reproduces both bilocal and FNN correlations in simulation. On the 156-qubit superconducting processor \textit{ibm\_kingston}, we observe bilocal nonlocality with $\mathcal{S}_{\rm BLHV}=1.067(6)>1$ after readout-error mitigation, while the FNN witnesses reach $99\%$ and $96\%$ of their certification thresholds, implying the substantially stronger requirements for FNN certification. We further show that Bob's joint measurement determines the accessible level of network nonlocality: bilocal and FNN certification are optimized by different measurement settings, while both violations can disappear even for maximally entangled states. These results identify measurement geometry as an independent resource for network nonlocality and provide a practical route toward its certification on programmable quantum processors.
\end{abstract}

\maketitle

\section{Introduction}

Quantum networks distribute entanglement among spatially separated parties and provide the foundation for quantum communication, cryptography, and distributed sensing \cite{Kimble2008, Wehner2018,https://doi.org/10.1002/lpor.202100219,RevModPhys.95.045006,Guo2020}. A central question is how to certify that the observed correlations cannot be explained by any classical model \cite{Brunner2014}. Unlike the standard Bell scenario, where nonlocality is determined by measurements on a single entangled source \cite{PhysicsPhysiqueFizika.1.195, PhysRevLett.23.880}, quantum networks contain multiple independent sources and therefore exhibit a richer structure of nonclassical correlations \cite{Fritz_2012, Tavakoli_2022}.

The simplest example is the bilocal Alice--Bob--Charlie network, in which two independent sources connect Bob to Alice and Charlie \cite{PhysRevA.85.032119}. The independence of the sources introduces new forms of nonlocality beyond those captured by Bell inequalities. Correlations may violate the bilocal hidden-variable (BLHV) model, which assumes two independent classical sources \cite{PhysRevA.85.032119}. More strongly, they may exhibit full network nonlocality (FNN), excluding every model in which at least one source remains classical \cite{PhysRevLett.128.010403}. Certifying FNN is therefore a substantially stronger requirement than demonstrating BLHV violation \cite{PRXQuantum.6.020317}.

A key feature of network nonlocality is that the measurement performed at the intermediate node can be as important as the entanglement distributed by the sources. In the Alice--Bob--Charlie network, Bob performs a joint measurement on particles received from two independent sources, effectively implementing entanglement swapping and establishing correlations between the outer parties. Consequently, the observed network correlations depend not only on the distributed entanglement but also on the structure of Bob's measurement \cite{Tavakoli_2022}.

This dependence is reflected directly in the BLHV and FNN witnesses, which are constructed from genuinely nonlocal observables involving multiple parties \cite{PhysRevA.85.032119, PhysRevLett.128.010403}. Such observables cannot, in general, be reduced to independent local measurements without altering the underlying network correlations. Nonlocal measurements therefore arise naturally in the certification of network nonlocality and constitute an intrinsic part of the underlying physics rather than merely a technical implementation detail \cite{PhysRevLett.116.070404,PhysRevLett.129.030502,PhysRevLett.126.220401,PhysRevResearch.7.013239}.

BLHV and FNN correlations are now well established both theoretically \cite{PhysRevA.85.032119, PhysRevLett.128.010403} and experimentally \cite{Carvacho2017, doi:10.1126/sciadv.1602743, Wang2023, PhysRevLett.130.190201, Hakansson2022}. Existing studies have primarily focused on demonstrating witness violations for specific measurement configurations. In contrast, much less attention has been paid to how the required nonlocal observables should be measured within a general quantum-network framework and, more fundamentally, to how the choice of nonlocal measurement influences the level of network nonlocality that can be certified.

In this work, we develop an operational framework based on ancilla-assisted meter circuits \cite{PhysRevResearch.7.013239} for evaluating the nonlocal observables in the BLHV and FNN witnesses. The framework reproduces the theoretical predictions in simulation and enables their realization on a superconducting quantum processor. While bilocal nonlocality can be certified on current hardware, the FNN witnesses approach but do not exceed their certification thresholds. This directly exposes the gap between bilocal and FNN certification and illustrates the greater experimental demands associated with certifying the stronger FNN correlations.

Beyond certification, we use the framework to investigate the role of measurement design itself. By continuously varying Bob's joint measurement operator, we show that the accessible level of nonclassicality is determined not only by the entanglement distributed by the sources but also by the geometry of the intermediate-node measurement. Remarkably, BLHV and FNN certification are optimized by different measurement settings: the corresponding violation regions are shifted relative to one another in parameter space and can disappear entirely even when the network is supplied with maximally entangled states. These results establish measurement geometry as an independent resource for network nonlocality.

The remainder of this paper is organized as follows. Section~\ref{sec:theory} introduces the hierarchy of network correlation sets and the corresponding witnesses. Section~\ref{sec:methods} presents the bilocal network and quantum circuit implementation. Section~\ref{sec:results} reports the numerical and hardware results. Finally, Sec.~\ref{sec:conclusion} summarizes our findings.

\section{Theoretical Framework}
\label{sec:theory}
Entangled quantum states can generate correlations that cannot be reproduced by classical hidden-variable theories. In the bipartite Bell scenario, this phenomenon is captured by the violation of a local hidden-variable (LHV) model \cite{PhysicsPhysiqueFizika.1.195}. Quantum networks generalize the Bell framework to multiple parties connected by independent sources, which are a key structural constraint \cite{Tavakoli_2022}. This additional constraint leads to a richer hierarchy of nonlocality and corresponding classical models \cite{Tavakoli_2022, PhysRevA.85.032119, Fritz_2012, PhysRevA.98.022113, PhysRevLett.123.070403, PRXQuantum.6.020317}.

\begin{figure}[t]
    \centering
    \includegraphics[width=\linewidth]{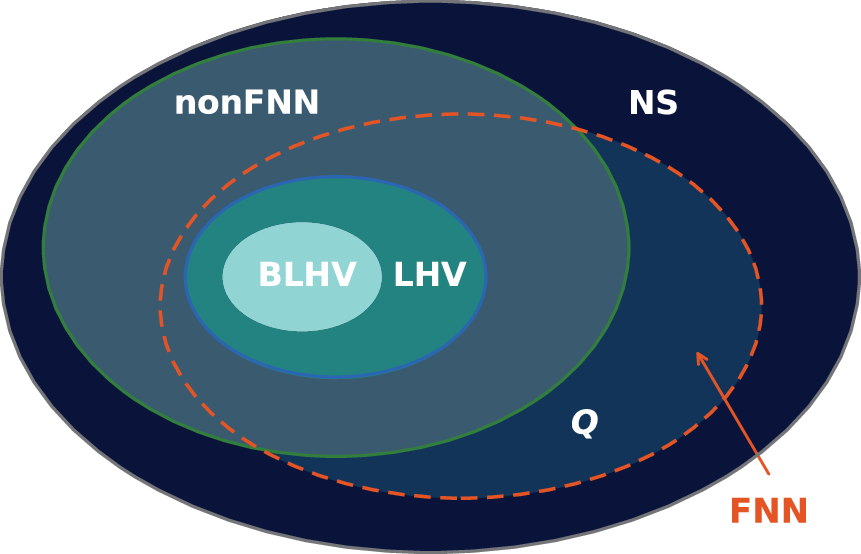}
\caption{\textbf{Hierarchy of correlation sets in quantum networks.}
The classical sets obey
$\mathrm{BLHV}\subset\mathrm{LHV}\subset\mathrm{nonFNN}\subset\mathrm{NS}$.
BLHV corresponds to independent classical sources, LHV allows shared classical randomness, and nonFNN allows one source to generate arbitrary NS correlations. The quantum set (Q) lies between LHV and NS but is incomparable with nonFNN. Violations of BLHV and nonFNN certify bilocal network nonlocality and full network nonlocality (FNN), respectively.}
    \label{fig1}
\end{figure}

All correlations in the network satisfy the no-signaling (NS) principle: changing the measurement setting of one party cannot alter the outcome statistics of any other party \cite{Popescu1994, PhysRevA.71.022101, Chaves2015}. Classical network models correspond to progressively smaller subsets of the NS set and form the hierarchy
\begin{equation}
\mathrm{BLHV} \subset \mathrm{LHV} \subset \mathrm{nonFNN} \subset \mathrm{NS},
\label{eq:hierarchy}
\end{equation}
shown in Fig.~\ref{fig1}. The BLHV model assumes two independent classical sources, whereas the LHV model allows arbitrary classical correlations between them. The nonFNN set is more permissive, requiring only that at least one source be classical while the other may generate arbitrary NS correlations \cite{PhysRevLett.128.010403}.

The quantum set (Q) consists of correlations generated by local measurements on shared entangled states, which satisfies
\(
\mathrm{LHV} \subset \mathrm{Q} \subset \mathrm{NS}.
\)
However, it is incomparable with nonFNN, where some quantum correlations lie outside nonFNN while nonFNN includes post-quantum NS resources. Violating one of these sets certifies the corresponding form of network nonlocality. In particular, a correlation is FNN if it lies outside nonFNN, implying that neither source can be replaced by a classical one, even when the other is allowed arbitrary NS power \cite{PhysRevLett.128.010403}.

\subsection{Local Hidden-Variable (LHV) Model}
The LHV model provides the standard classical description of Bell correlations \cite{PhysicsPhysiqueFizika.1.195}. For two parties, Alice and Charlie, with measurement inputs $x$ and $z$ and outcomes $a$ and $c$, the correlations admit the decomposition
\begin{equation}
    P(a,c|x,z)
    =
    \int d\lambda\,\rho(\lambda)\,
    P(a|x,\lambda)\,
    P(c|z,\lambda),
    \label{eq:LHV}
\end{equation}
where $\lambda$ is a shared hidden variable distributed according to $\rho(\lambda)$. Correlations of this form satisfy all Bell inequalities, including the Clauser-Horne-Shimony-Holt (CHSH) inequality \cite{PhysRevLett.23.880}. Consequently, any Bell-inequality violation rules out an LHV description.

\subsection{Bilocal Hidden-Variable (BLHV) Model}
The LHV model does not capture the causal structure of quantum networks, where correlations are generated by multiple independent sources \cite{doi:10.1126/sciadv.aea8571}. In the bilocal network, two independent sources distribute hidden variables $\lambda_1$ and $\lambda_2$ to the pairs (Alice, Bob) and (Bob, Charlie), respectively. The resulting correlations take the form
\begin{align}
P(a,b,c|x,y,z)
&=
\iint d\lambda_1\, d\lambda_2\,
\rho_1(\lambda_1)\rho_2(\lambda_2)
\nonumber\\
&\quad\times
P(a|x,\lambda_1)\,
P(b|y,\lambda_1,\lambda_2)\,
P(c|z,\lambda_2),
\label{eq:BLHV}
\end{align}
where source independence is encoded by the factorization
$\rho(\lambda_1,\lambda_2)=\rho_1(\lambda_1)\rho_2(\lambda_2)$.

Correlations compatible with Eq.~(\ref{eq:BLHV}) satisfy the nonlinear bilocal inequality \cite{PhysRevA.85.032119}
\begin{equation}
\mathcal{S}_{\rm BLHV}
=
\sqrt{|\mathcal{I}|}
+
\sqrt{|\mathcal{J}|}
\le 1,
\label{eq:bilocal_ineq_general}
\end{equation}
where $\mathcal{I}$ and $\mathcal{J}$ are linear combinations of the observed probabilities. A violation of this inequality certifies network nonlocality beyond any model with independent classical sources. The explicit expressions for $\mathcal{I}$ and $\mathcal{J}$ are given in Sec.~\ref{sec:methods}.

\subsection{Full Network Nonlocality (FNN)}
The FNN is the strongest form of nonclassicality in the hierarchy of Fig.~\ref{fig1} \cite{PhysRevLett.128.010403}. A correlation is FNN if it lies outside the nonFNN set, i.e., if it cannot be reproduced by any model in which at least one source remains classical.

In the bilocal network, one family of such hybrid models assumes that the Alice--Bob source is classical, while the Bob--Charlie source may generate arbitrary NS correlations:
\begin{equation}
P(a,b,c|x,y,z)
=
\int d\lambda\,\rho(\lambda)\,
P(a|x,\lambda)\,
P(b,c|y,z,\lambda),
\label{eq:FNN}
\end{equation}
where $P(b,c|y,z,\lambda)$ is an arbitrary NS distribution \cite{Hakansson2022}. The second family is obtained by exchanging the roles of the two sources,
$P(a,b|x,y,\lambda)P(c|z,\lambda)$.
The union of these two families defines the nonFNN set.

To certify FNN, both families must be excluded simultaneously. In this work, this is achieved using the witnesses $\mathcal{R}_{\mathrm{CNS}}$ and $\mathcal{R}_{\mathrm{NSC}}$, which bound the classical--NS and NS--classical models, respectively (see Sec.~\ref{sec:methods}).
Correlations that violate both hybrid models lie outside nonFNN and therefore certify FNN.


\section{Bilocal network and circuit implementation}
\label{sec:methods}

\begin{figure}[b]
    \centering
    \includegraphics[width=\linewidth]{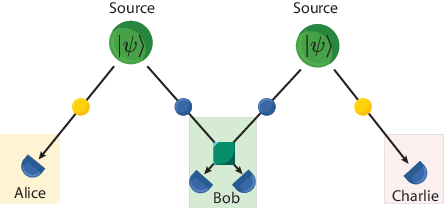}
    \caption{\textbf{Bilocal network configuration.} Two independent entangled sources each distribute one qubit to Bob and one to an outer party, generating pairs $(A, B_A)$ and $(B_C, C)$. Bob holds both subsystems $B_A$ and $B_C$ and performs joint measurements, mediating correlations between Alice and Charlie.}
    \label{fig2}
\end{figure}

\subsection{Bilocal network}
We consider a bilocal network shown in Fig.~\ref{fig2}, consisting of Alice, Bob, and Charlie. Bob holds two subsystems, $B_A$ and $B_C$, shared with Alice and Charlie, respectively, such that $B=B_A\otimes B_C$. By performing a joint measurement on $B_A$ and $B_C$, Bob implements entanglement swapping and thereby generates correlations between Alice and Charlie. The nature of these correlations depends on the measurement performed by Bob and determines the level of network nonlocality that can be observed.

The network is initialized in two identical and independent entangled pairs,
\begin{equation}
|\Psi(\theta)\rangle
=
|\psi(\theta)\rangle_{A,B_A}
\otimes
|\psi(\theta)\rangle_{B_C,C},
\label{eq:Psi}
\end{equation}
where
\begin{equation}
|\psi(\theta)\rangle
=
\cos\!\left(\frac{\theta}{2}\right)|01\rangle
+
\sin\!\left(\frac{\theta}{2}\right)|10\rangle.
\label{eq:state}
\end{equation}
The parameter $\theta$ tunes the state from a separable ($\theta=0,\pi$) to a maximally entangle ($\theta=\pi/2$), enabling a systematic study of network nonlocality across different entanglement regimes \cite{PhysRevResearch.7.013239}.

We choose the measurement settings as follow.
Alice measures $A_0 = X$ or $A_1 = Z$; Bob measures $B_0 = Z \otimes Z$ or $B_1 = X \otimes X$; Charlie measures $C_0 = (Z + X)/\sqrt{2}$ or $C_1 = (Z - X)/\sqrt{2}$, 
where $X$, $Y$, and $Z$ are Pauli operators. This configuration is known to produce a strong violation of the bilocal inequality \cite{PhysRevA.85.032119}.
For binary inputs $x,y,z\in\{0,1\}$ and outputs $a,b,c\in\{0,1\}$, the bilocal witness is constructed from
\begin{align}
\mathcal{I}
&=
\frac{1}{4}
\sum_{x,z}
\langle A_x B_0 C_z\rangle,
\label{eq:I}
\\
\mathcal{J}
&=
\frac{1}{4}
\sum_{x,z}
(-1)^{x+z}
\langle A_x B_1 C_z\rangle,
\label{eq:J}
\end{align}
where
\begin{equation}
\langle A_x B_y C_z\rangle
=
\sum_{a,b,c}
(-1)^{a+b+c}
P(a,b,c|x,y,z).
\end{equation}
These quantities correspond to $\mathcal{I}_{22}$ and $\mathcal{J}_{22}$ of Ref.~\cite{PhysRevA.85.032119}. The BLHV model imposes the bound
\begin{equation}
\mathcal{S}_{\rm BLHV}
=
\sqrt{|\mathcal{I}|}
+
\sqrt{|\mathcal{J}|}
\le 1.
\label{eq:bilocal_ineq}
\end{equation}

\begin{figure*}[ht] 
    \centering
    \includegraphics[width=0.8\textwidth]{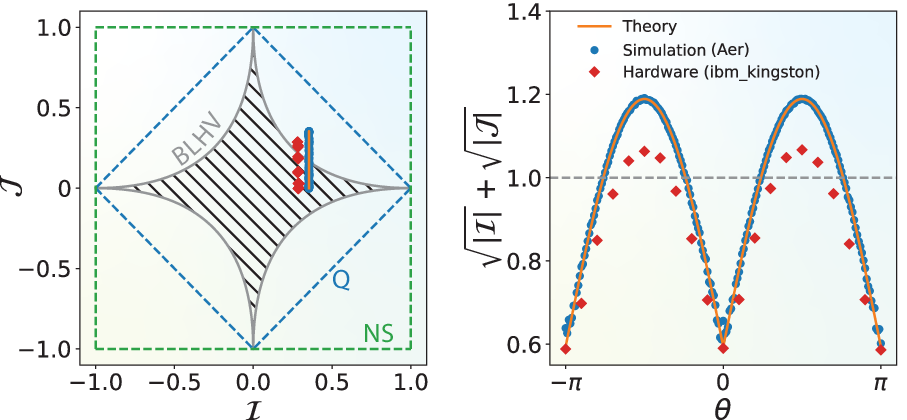} 
    \caption{\textbf{Analysis of the BLHV model on bilocal network.} (Left) Theoretical, simulation (\textit{qiskit\_aer}), and hardware (\textit{ibm\_kingston}) results for $\mathcal{I}$ and $\mathcal{J}$ projected onto the $\mathcal{I}$--$\mathcal{J}$ plane, obtained by sweeping $\theta$ over the full range $[-\pi,\pi]$. The hatched star-shaped region represents the set of correlations compatible with the BLHV model, bounded by $\sqrt{|\mathcal{I}|} + \sqrt{|\mathcal{J}|} \leq 1$. The blue dashed diamond and green dashed square indicate the quantum (Q) and no-signaling (NS) boundaries, respectively. The simulation data trace a closed curve lying outside the bilocal region, in agreement with the theoretical predictions. (Right) The bilocal quantity $\sqrt{|\mathcal{I}|} + \sqrt{|\mathcal{J}|}$ as a function of $\theta$. The dashed horizontal line marks the classical bound of $1$, which is exceeded in broad regions centered around $\theta \approx \pm\pi/2$, reaching a maximum of approximately $1.2$ at maximal entanglement. 
Red diamonds show the results from \textit{ibm\_kingston} quantum processor with $32\,768$ shots per circuit; statistical error bars are smaller than the symbol size.
    }
    \label{fig3}
\end{figure*}

Correlations beyond the nonFNN set are certified using the network witnesses \cite{PhysRevLett.128.010403,Hakansson2022}
\begin{align}
\mathcal{R}_{\mathrm{CNS}} &\coloneqq 2\langle A_0 \bar{B}_1 (C_0 - C_1) \rangle + \langle A_1 B_0 (2C_0 + C_1) \rangle - \langle B_0 \rangle \nonumber \\
&\quad + \bigl(\langle A_1 B_0 \rangle + \langle B_0 C_0 \rangle - \langle C_0 \rangle\bigr)\langle C_1 \rangle,
\label{eq:R_CNS}
\\
\mathcal{R}_{\mathrm{NSC}} &\coloneqq 2\langle A_0 \bar{B}_1 (C_0 - C_1) \rangle + \langle A_1 B_0 (C_0 + 2C_1) \rangle - \langle B_0 \rangle \nonumber \\
&\quad + \langle A_1 \rangle\bigl(\langle A_1 B_0 \rangle + \langle B_0 C_1 \rangle + \langle C_0 - C_1 - A_1 \rangle\bigr),
\label{eq:R_NSC}
\end{align}
where
\begin{equation}
\bar{B}_1
=
|\phi^+\rangle\langle\phi^+|
-
|\phi^-\rangle\langle\phi^-|
=
\frac{1}{2}\left(X\otimes X-Y\otimes Y\right)
\label{eq:B1bar}
\end{equation}
is a partial Bell-state measurement on Bob's qubits \cite{PhysRevLett.128.010403,Hakansson2022}. Both witnesses then have a classical upper bound of $3$, and their simultaneous violation,
\begin{equation}
\mathcal{R}_{\mathrm{CNS}} > 3
\quad\text{and}\quad
\mathcal{R}_{\mathrm{NSC}} > 3,
\label{eq:FNN_condition}
\end{equation}
exclude both classical--NS and NS--classical models, thereby certifying FNN. The asymmetry between $\mathcal{R}_{\mathrm{CNS}}$ and $\mathcal{R}_{\mathrm{NSC}}$ reflects the directed network structure, while the involvement of single-, two-, and three-body correlators makes FNN certification strictly stronger than a violation of the BLHV inequality.

\subsection{Quantum Circuit Implementation}
The bilocal network is implemented on four qubits, representing Alice, Bob's two subsystems $(B_A,B_C)$, and Charlie, respectively (see Fig.~\ref{fig6} in Appendix~\ref{appA}). The circuit consists of state-preparation qubits and  ancilla-assisted measurement qubits. The state-preparation part independently generates the two entangled pairs of Eq.~\eqref{eq:Psi}, thereby preparing the network state $|\Psi(\theta)\rangle$.

To evaluate the BLHV and FNN witnesses, we measure a set of one-, two-, and three-body correlators involving Alice, Bob, and Charlie. Since the FNN witnesses contain nonlocal observables acting jointly on Bob's two qubits, the corresponding expectation values are extracted using ancilla-assisted circuits that preserve the network state during the measurement process \cite{PhysRevResearch.7.013239}.

The measured correlators include
\begin{gather*}
\langle A_1 \rangle,\ \langle B_0 \rangle,\ \langle C_z \rangle,\ \langle A_1 B_0 \rangle,\ \langle B_0 C_z \rangle,\ \langle A_x B_y C_z \rangle,\\
\langle A_0 \, (B_0 B_1) \, C_z \rangle,
\end{gather*}
with $x,y,z\in\{0,1\}$. These quantities are combined according to Eqs.~\eqref{eq:I}--\eqref{eq:R_NSC} to evaluate the BLHV and FNN witnesses. Detailed circuit constructions and measurement protocols are provided in Appendix~\ref{appA}.

\section{Results}
\label{sec:results}
We implement the bilocal network described in Sec.~\ref{sec:methods} on both a simulator and quantum hardware. Simulator results are obtained from the probability distribution using the Aer simulator (\textit{qiskit\_aer}) with $10^5$ shots per circuit, allowing direct verification of the circuit implementation while isolating finite-sampling effects. Hardware results are obtained on the 156-qubit superconducting processor \textit{ibm\_kingston} (IBM Heron architecture). The circuits are executed as a single parametrized job for $21$ values of $\theta\in[-\pi,\pi]$, with $32\,768$ shots per circuit. We employ transpilation at optimization level 3, dynamical decoupling, measurement twirling, and readout-error mitigation based on single-qubit calibration data (mean readout error $1.3\%$). Reported uncertainties account for shot-noise propagation.

We focus on three key aspects: violation of the BLHV model, certification of FNN correlations, and the measurement selectivity of network nonlocality.

\subsection{Violation of the BLHV inequality}
We evaluate the bilocal correlators $\mathcal{I}$ and $\mathcal{J}$ defined in Eqs.~\eqref{eq:I} and \eqref{eq:J} as functions of the state parameter $\theta$ and test the BLHV inequality~\eqref{eq:bilocal_ineq}.

Figure~\ref{fig3}(left) shows the results of $\mathcal{I}$ and $\mathcal{J}$ projected onto the $\mathcal{I}$-$\mathcal{J}$ plane, swept over $\theta \in [-\pi, \pi]$. The shaded star-shaped region represents correlations compatible with the BLHV model, bounded by $\mathcal{S}_{\rm BLHV} \leq 1$. The simulation data trace a closed curve lying clearly outside the bilocal region over a broad range of $\theta$, in excellent agreement with theoretical predictions, demonstrating robust violation of the BLHV inequality. The hardware results (red diamonds) exhibit systematic deviations from the theoretical curves due to the combined effects of gate errors, decoherence, and measurement imperfections.

Figure~\ref{fig3}(right) shows $\mathcal{S}_{\rm BLHV} = \sqrt{|\mathcal{I}|} + \sqrt{|\mathcal{J}|}$ as a function of $\theta$. The classical bound is exceeded over broad parameter regions centered around $\theta = \pm\pi/2$, reaching a maximum of approximately $1.2$ at maximal entanglement, and is weakest away from these regions, consistent with reduced entanglement. The agreement between simulation and theory across the full parameter range confirms that the nonlocal measurement method faithfully implements the target state and measurement configuration.

The hardware results follow the theoretical curve closely, with the correlator $\mathcal{I}$ attenuated from $1/(2\sqrt{2})\approx 0.354$ to $\approx 0.28$, approximately independently of $\theta$. Crucially, the square-root structure of $\mathcal{S}_{\mathrm{BLHV}}$ compresses this attenuation, and the bilocal bound is still violated on hardware: $\mathcal{S}_{\mathrm{BLHV}} = 1.067 \pm 0.006$ at $\theta = \pm\pi/2$, exceeding the classical bound by more than six standard deviations, with violation sustained over the window $|\theta| \in [0.4\pi, 0.6\pi]$.

These results establish robust bilocal nonlocality over a wide entanglement range. As shown in the following subsection, however, this robustness does not extend straightforwardly to the FNN correlations.

\subsection{Certification of FNN correlations}
We evaluate the FNN witnesses $\mathcal{R}_{\mathrm{CNS}}$ and $\mathcal{R}_{\mathrm{NSC}}$ defined in Eqs.~\eqref{eq:R_CNS} and \eqref{eq:R_NSC} and test the simultaneous violation condition~\eqref{eq:FNN_condition}.

\begin{figure}[t]
    \centering
    \includegraphics[width=\linewidth]{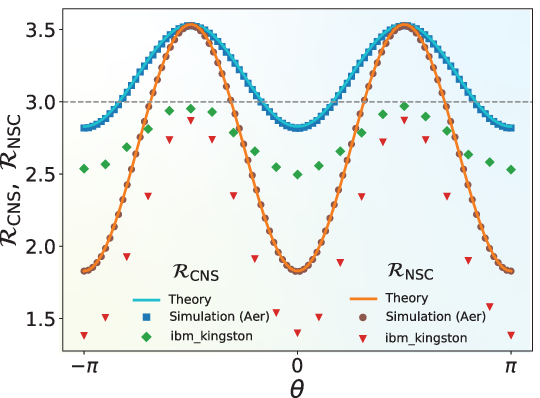}
    \caption{\textbf{FNN witnesses $\mathcal{R}_{\mathrm{CNS}}$ and $\mathcal{R}_{\mathrm{NSC}}$ as functions of the state parameter $\theta$.} Theoretical, simulation (\textit{qiskit\_aer}), and hardware processor (\textit{ibm\_kingston}) values of $\mathcal{R}_{\mathrm{CNS}}$ and $\mathcal{R}_{\mathrm{NSC}}$ are shown as functions of $\theta$. The dashed horizontal line marks the classical bound of $3$, above which FNN correlations are certified. Both witnesses reach the maximum value $5/\sqrt{2} \approx 3.54$ near $\theta = \pm\pi/2$, corresponding to maximal entanglement. Hardware values are readout-error mitigated with $32\,768$ shots per circuit, and statistical error bars are smaller than the symbol size.
    }
    \label{fig4}
\end{figure}

Figure~\ref{fig4} shows the results of $\mathcal{R}_{\mathrm{CNS}}$ and $\mathcal{R}_{\mathrm{NSC}}$ as functions of $\theta$, with the classical bound of $3$ indicated by the dashed horizontal line. Both witnesses exceed this bound in regions centered around $\theta = \pm\pi/2$, reaching the maximum value $5/\sqrt{2} \approx 3.54$ at maximal entanglement, in agreement with the quantum prediction of Refs.~\cite{PhysRevLett.128.010403, Hakansson2022}, and simultaneous violation is sustained for $|\theta| \in [0.31\pi, 0.68\pi]$. The two witnesses behave nearly identically as functions of $\theta$ at maximal entanglement, reflecting their symmetric construction under exchange of Alice and Charlie. The agreement between simulation and theory across the full $\theta$ range again confirms that the nonlocal measurement circuits successfully access the joint multipartite correlators without destroying the entanglement required for FNN certification.

On hardware, both witnesses follow the $\theta$ dependence of the theoretical prediction with an overall noise-induced attenuation, reaching $\mathcal{R}_{\mathrm{CNS}} = 2.97 \pm 0.02$ and $\mathcal{R}_{\mathrm{NSC}} = 2.87 \pm 0.02$ at $\theta = \pi/2$, i.e., $99\%$ and $96\%$ of the classical bound. At the present noise level, the FNN witnesses therefore remain just below the certification threshold. In contrast, the bilocal violation survives comfortably. This gap indicates how much harder it is to observe FNN than BLHV violation on current hardware. The shortfall of only a few percent indicates that FNN certification on superconducting processors is within reach of modest improvements in gate fidelity or error mitigation.

\begin{figure*}[t]
    \centering
    \includegraphics[width=0.8\linewidth]{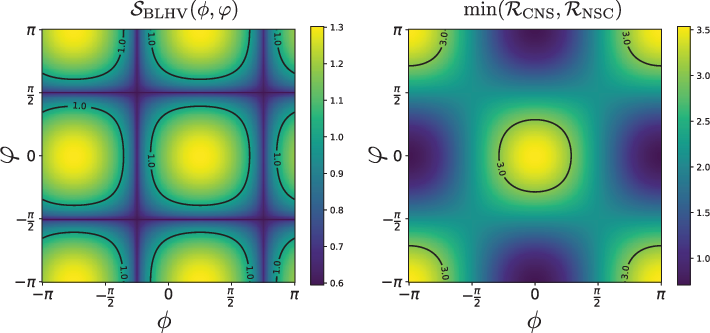}
    \caption{\textbf{Dependence of BLHV and FNN witnesses on the measurement parameters $\phi$ and $\varphi$.} (Left) Distribution of $\mathcal{S}_{\mathrm{BLHV}}(\phi,\varphi) = \sqrt{|\mathcal{I}(\phi,\varphi)|} + \sqrt{|\mathcal{J}(\phi,\varphi)|}$ over the $(\phi,\varphi)$ plane, with the black contour marking the classical bound $\mathcal{S}_{\mathrm{BLHV}} = 1$. The distribution exhibits a periodic structure with local maxima at $(\phi,\varphi) = (\pi/4 + k\pi, n\pi)$ for $k, n \in \mathbb{Z}$. (Right) Distribution of $\min(\mathcal{R}_{\mathrm{CNS}}, \mathcal{R}_{\mathrm{NSC}})$, evaluated with the partial-BSM observable $\bar{B}_1(\phi,\varphi)$, over the same parameter space, with the black contour at the classical bound $3$. The FNN distribution reaches its maxima at $(\phi,\varphi) = (k\pi, n\pi)$ with $k + n \in 2\mathbb{Z}$. The BLHV violation region covers about $53\%$ of the parameter space, while the FNN region covers about $14\%$; the two sets of maxima are shifted relative to each other by $\pi/4$ in $\phi$, establishing that BLHV violation is a necessary but not sufficient condition for FNN certification.}
    \label{fig5}
\end{figure*}

Comparing Figs.~\ref{fig3} and \ref{fig4}, the FNN violations are confined to a narrower range of $\theta$ ($|\theta| \in [0.31\pi, 0.68\pi]$) than the BLHV violations ($|\theta| \in [0.24\pi, 0.76\pi]$), as expected for a strictly stronger notion of nonclassicality: ruling out models in which one source retains arbitrary NS power requires more entanglement than ruling out independent classical sources. The relative violation is correspondingly smaller, $18\%$ above the FNN bound at maximal entanglement compared with $19\%$ for $\mathcal{S}_{\rm BLHV}$, but is concentrated in fewer correlators, which makes the FNN witnesses more fragile to noise, as observed on hardware.

These results demonstrate that the quantum circuit with fixed measurement settings certifies correlations beyond all network models with at least one classical source. The distinct violation ranges of the two witnesses further suggest that network nonlocality certification depends not only on entanglement strength, but also on the specific structure of the joint measurements at Bob's node, a question addressed systematically in the following subsection.

\subsection{Measurement selectivity of network nonlocality}
We investigate how network nonlocality depends on the structure of joint measurements at the intermediate node. In the previous subsections, Bob's settings were fixed to $B_0 = Z \otimes Z$, $B_1 = X \otimes X$, and $\bar{B}_1$ of Eq.~\eqref{eq:B1bar}. These settings violate both the BLHV and FNN bounds. We now relax this choice and examine whether such violations are generic or measurement-specific.

Bob's joint measurement is parametrized as
\begin{equation}
B_1(\phi,\varphi) = (\cos\phi\,X + \sin\phi\,Z)\otimes(\cos\varphi\,X + \sin\varphi\,Z), \label{eq:Bob_param}
\end{equation}
which yields dichotomic outcomes $\pm 1$ for all $\phi$ and $\varphi$. The state is fixed at $\theta = \pi/2$ and the settings of Alice and Charlie are unchanged. Operationally, $B_1(\phi,\varphi)$ is implemented by local rotations $R_y(-\phi)$ and $R_y(-\varphi)$ on Bob's two qubits followed by a $Z \otimes Z$ measurement. The FNN witnesses use the rotated partial-BSM observable $\bar{B}_1(\phi,\varphi) = [B_1(\phi,\varphi) - Y \otimes Y]/2$. This form follows from Eq.~\eqref{eq:B1bar} because the rotations leave the $Y \otimes Y$ component unchanged. The BLHV and FNN witnesses are then evaluated over $(\phi,\varphi) \in [-\pi,\pi]$.

Figure~\ref{fig5} shows the distributions of $\mathcal{S}_{\mathrm{BLHV}}(\phi,\varphi)$ and $\min(R_{\mathrm{CNS}}, R_{\mathrm{NSC}})$ over the $(\phi,\varphi)$ plane, with black contours marking the classical bounds of $1$ and $3$, respectively. The distribution of $\mathcal{S}_{\mathrm{BLHV}}$ exhibits a periodic structure with local maxima at $(\phi,\varphi) = (\pi/4 + k\pi, n\pi)$ for $k, n \in \mathbb{Z}$, while the FNN distribution reaches its maxima at $(\phi,\varphi) = (k\pi, n\pi)$ with $k + n \in 2\mathbb{Z}$. The BLHV violation region covers about $53\%$ of the parameter space, while the FNN certification region is markedly smaller, about $14\%$, and the two sets of maxima are shifted relative to each other by $\pi/4$ in $\phi$ while remaining aligned in $\varphi$. Large portions of the parameter space exhibiting BLHV violation therefore fall outside the FNN certification region, directly establishing that BLHV violation is a necessary but not sufficient condition for FNN certification.


From the perspective of entanglement swapping, this shift reflects the different demands the two tests place on Bob's measurement. BLHV violation only requires the swapping operation to transfer sufficient three-body correlations, and a broad family of orientations achieves this. FNN certification also constrains the single- and two-body correlators entering $\mathcal{R}_{\mathrm{CNS}}$ and $\mathcal{R}_{\mathrm{NSC}}$, and only the narrower class of measurements around $(k\pi, n\pi)$ satisfies all of them.

These results have a direct practical implication. Both violations can fail even when each source produces a maximally entangled pair. Certification therefore depends on the measurement configuration as much as on entanglement. The $\pi/4$ shift also means that a small deviation in Bob's orientation can move the system out of the FNN regime while remaining within the BLHV region. The map in Fig.~\ref{fig5} provides an operational guide for targeting a specific level of network nonlocality.

\section{Conclusion}
\label{sec:conclusion}

We have developed an operational framework for certifying network nonlocality in the bilocal Alice--Bob--Charlie network using ancilla-assisted meter circuits. The framework enables direct evaluation of the nonlocal observables entering the BLHV and FNN witnesses and provides a practical route for implementing network-nonlocality tests on programmable quantum processors.

The framework reproduces the expected behavior of both BLHV and FNN witnesses in simulation and enables their experimental investigation on the 156-qubit superconducting processor \textit{ibm\_kingston}. Bilocal nonlocality was successfully certified after readout-error mitigation, while the FNN witnesses approached, but did not exceed, their certification thresholds. This highlights the substantially stronger requirements associated with FNN certification compared with BLHV violation.

Beyond certification, we used the framework to investigate the role of measurement design in the bilocal network. By continuously varying Bob's joint measurement, we found that the accessible level of network nonlocality depends not only on the entanglement distributed by the sources but also on the geometry of the intermediate-node measurement itself. In particular, BLHV and FNN violations can disappear even when the network is supplied with maximally entangled states, and their optimal regions occur at different measurement settings. These results identify measurement geometry as an independent resource for network nonlocality, alongside entanglement.

Our results establish a direct connection between nonlocal measurements, network-nonlocality certification, and programmable quantum hardware. The operational framework introduced here can be extended to larger network topologies, more general causal structures, and advanced quantum-network protocols, providing a versatile platform for exploring nonclassical correlations in future quantum networks.

\section*{Acknowledgments}
This work was supported by the Tohoku Initiative for Fostering Global Researchers for Interdisciplinary Sciences (TI-FRIS) under MEXT's Strategic Professional Development Program for Young Researchers. We acknowledge the use of IBM Quantum services for this work. The views expressed are those of the authors and do not reflect the official policy or position of IBM or the IBM Quantum team.

\appendix
\section{Quantum circuit implementation}\label{appA}
The following circuits were constructed using Qiskit to evaluate the expectation values of the observables.

\begin{figure}[b]
    \centering
    \includegraphics[width=0.75\linewidth]{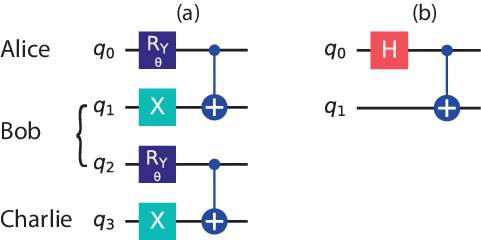}
    \caption{
    \textbf{(a) System circuit (State preparation).} It consists of two 
    identical subcircuits, each preparing the state $|\phi(\theta)\rangle$. 
    In each subcircuit, an $R_y(\theta)$ gate is applied to the first qubit 
    and a Pauli-$X$ gate to the second, followed by a CNOT gate to generate 
    the Bell state $|\phi(\theta)\rangle$. \textbf{(b) Meter circuit.} It consists of a Hadamard gate 
    on the first qubit followed by a CNOT gate. The meter circuit is used to 
    perform nonlocal measurements on the system register.
    }
    \label{fig6}
\end{figure}

\subsection{State preparation (system) circuit}
To prepare the state $|\Psi(\theta)\rangle$ of Eq.~\eqref{eq:Psi} we use a circuit of four qubits $q_0$-$q_3$, where qubits $(q_0, q_1)$ represent the $(A, B_A)$ pair and qubits $(q_2, q_3)$ represent the $(B_C, C)$ pair. An identical gate sequence is applied independently to each pair, as illustrated in Fig.~\ref{fig6}(a). For the pair $(q_0, q_1)$, the sequence proceeds as follows. First, an $R_y(\theta) = \begin{pmatrix} \cos\frac{\theta}{2} & -\sin\frac{\theta}{2} \\ \sin\frac{\theta}{2} & \cos\frac{\theta}{2} \end{pmatrix}$ rotation gate is applied to $q_0$ and a Pauli-$X$ gate is applied to $q_1$, initializing the pair to
    \begin{align}
        |\psi_1\rangle &= \big(R_y(\theta) \otimes X\big)|00\rangle \nonumber \\
        &= \big( \cos\frac{\theta}{2}|0\rangle + \sin\frac{\theta}{2}|1\rangle \big) \otimes |1\rangle \nonumber \\
        &= \cos\frac{\theta}{2}|01\rangle + \sin\frac{\theta}{2}|11\rangle.
    \end{align}
Second, a CNOT gate with $q_0$ as control and $q_1$ as target entangles the pair, yielding
\begin{equation}
    |\psi(\theta)\rangle = 
    \cos\frac{\theta}{2}|01\rangle + 
    \sin\frac{\theta}{2}|10\rangle,
\end{equation}
in agreement with Eq.~\eqref{eq:state}. The same sequence applied to $(q_2, q_3)$ produces the second entangled pair, and the overall four-qubit state is the tensor product $|\Psi(\theta)\rangle$ as defined in Eq.~\eqref{eq:Psi}.

In the circuit, $q_0$ is Alice's qubit, $q_1$ and $q_2$ are Bob's two qubits, and $q_3$ is Charlie's qubit.

\begin{figure*}[t]
    \includegraphics[width=0.7\linewidth]{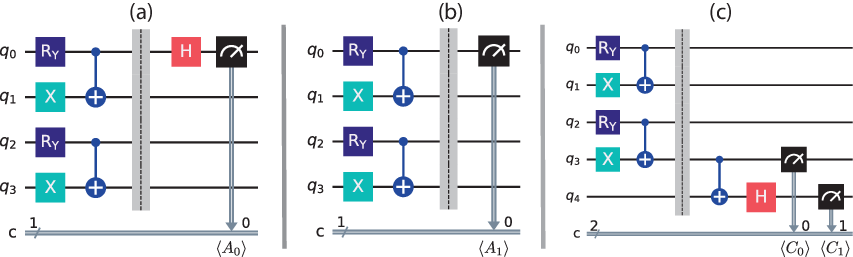}
    \caption{(a) Circuit for measuring $\langle A_0 \rangle 
    = \langle X \rangle$. 
    (b) Circuit for measuring $\langle A_1 \rangle 
    = \langle Z \rangle$.
    (c) Circuit for measuring $\langle C_0 \rangle 
    = \langle Z+X \rangle/\sqrt{2}$ and $\langle C_1 \rangle 
    = \langle Z-X \rangle/\sqrt{2}$.
    }
    \label{fig7}
\end{figure*}

\subsection{Meter circuit}\label{appA2}
Following the system circuit for state preparation, meter circuits are applied 
to perform nonlocal measurements on the system register. The meter circuit, 
illustrated in Fig.~\ref{fig6}(b), enables the extraction of expectation 
values without destroying the entanglement of the system. This is particularly 
important for the two- and three-party observables, which 
involve genuinely joint measurements across different nodes. The specific 
construction of each measurement circuit, including the configurations for 
single-party, two-party, and three-party expectation values, is described in 
detail in the following.

\subsection{Circuits for measurement expectation values}
\label{appA3}

Based on the definitions in Eqs.~\eqref{eq:I}, \eqref{eq:J}, \eqref{eq:R_CNS}, and \eqref{eq:R_NSC}, the evaluation of the BLHV and FNN witnesses requires the following expectation values:
\begin{gather*}
    \langle A_x \rangle,\
    \langle B_0 \rangle,\
    \langle C_z \rangle,\
    \langle A_1 B_0 \rangle,\
    \langle B_0 C_z \rangle,\
    \langle A_x B_y C_z \rangle,\\
    \langle A_0 \, (B_0 B_1) \, C_z \rangle,
\end{gather*}
for $x, y, z \in \{0,1\}$. Each expectation value is obtained from a dedicated circuit consisting of the system circuit followed by an appropriate measurement circuit, as described in Sec.~\ref{appA2}. The measurement strategy follows the nonlocal framework \cite{PhysRevResearch.7.013239}: single-qubit Pauli observables are read out projectively after an appropriate basis rotation, while observables involving Charlie's two settings or several parties are read out through ancilla-assisted CNOT interactions.

All classical post-processing takes a common form. Let $p_{\mathbf{m}}$ denote the probability of obtaining the classical outcome string $\mathbf{m} = m_{n-1}\cdots m_1 m_0$, where $m_r \in \{0,1\}$ is the bit recorded in classical register $c_r$. For a subset $\mathcal{S}$ of the classical registers we define the marginal parity
\begin{equation}
    E[\mathcal{S}] = \sum_{\mathbf{m}} (-1)^{\oplus_{r \in \mathcal{S}} m_r}\, p_{\mathbf{m}},
    \label{eq:parity}
\end{equation}
where $\oplus$ denotes addition modulo 2 and all registers outside $\mathcal{S}$ are summed over. Every expectation value below is either a single marginal parity or a linear combination of two marginal parities evaluated on the same shots, which preserves the shot-by-shot correlations between registers. We describe each case below.

\begin{figure*}[t]
        \includegraphics[width=0.8\linewidth]{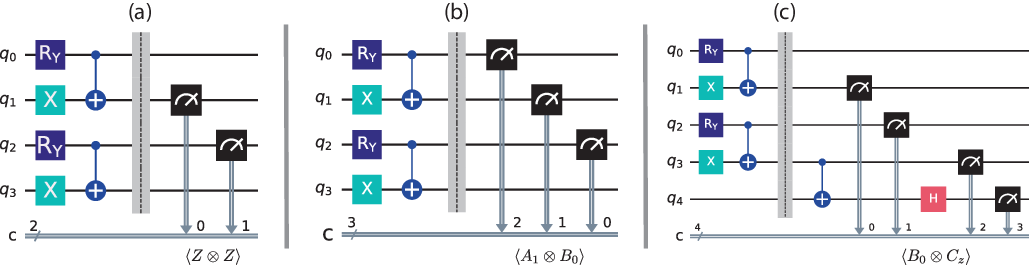}
        \caption{(a) Circuit for measuring $\langle B_0\rangle$.
    (b) Circuit for measuring $\langle A_1 B_0 \rangle$. 
    (c) Circuit for measuring $\langle B_0 C_z \rangle$.}
        \label{fig8}
\end{figure*}

\subsubsection{Single-party expectation values}

The simplest measurements are the single-party expectation values $\langle A_x \rangle$, obtained by direct projective measurement of Alice's qubit in the appropriate basis.

For $\langle A_0 \rangle = \langle X \rangle$, a Hadamard gate is applied to qubit $q_0$ before measurement to rotate from the $\sigma_x$ eigenbasis to the computational $Z$ basis, as illustrated in Fig.~\ref{fig7}(a). With the outcome recorded in register $c_0$, the expectation value is
\begin{equation}
    \langle A_0 \rangle = E[\{c_0\}] = p_0 - p_1,
\end{equation}
where $p_0$ and $p_1$ are the probabilities of obtaining outcomes $|0\rangle$ and $|1\rangle$, respectively.

For $\langle A_1 \rangle = \langle Z \rangle$, no basis rotation is required. Qubit $q_0$ is measured directly in the computational basis, as shown in Fig.~\ref{fig7}(b), and the expectation value is computed in the same way as above.

For $\langle C_0 \rangle$ and $\langle C_1 \rangle$, both settings are obtained from a single circuit. We introduce an ancilla qubit $q_4$ and apply a CNOT gate with $q_3$ as control and $q_4$ as target, followed by a Hadamard gate on $q_4$. Qubit $q_3$ is then measured into register $c_0$ and $q_4$ into register $c_1$, as illustrated in Fig.~\ref{fig7}(c). The two registers provide the $Z$-basis and $X$-basis readouts of Charlie's qubit,
\begin{equation}
    \langle Z_C \rangle = E[\{c_0\}], \qquad
    \langle X_C \rangle = E[\{c_1\}],
\end{equation}
and both measurement settings follow from the decomposition $C_z = [Z + (-1)^z X]/\sqrt{2}$ as
\begin{equation}
    \langle C_z \rangle = \frac{\langle Z_C \rangle + (-1)^z \langle X_C \rangle}{\sqrt{2}}, \quad z = 0, 1.
    \label{eq:Cz_combination}
\end{equation}
This ancilla-assisted scheme evaluates both settings from a single circuit execution.

To measure $\langle B_0 \rangle = \langle Z \otimes Z
\rangle$, we measure qubits $q_1$ and $q_2$ directly
in the computational basis into registers $c_0$ and
$c_1$, as shown in Fig.~\ref{fig8}(a). The measurement
outcomes are $p_{ij}$ for $ij \in \{00, 01, 10, 11\}$,
and the expectation value is computed as
\begin{equation}
    \langle B_0 \rangle = E[\{c_0, c_1\}] = p_{00} + p_{11} - p_{01} - p_{10},
\end{equation}
where the sign of each term reflects the eigenvalue
$(-1)^{i+j}$ of the $Z \otimes Z$ operator for the
corresponding two-qubit outcome $|ij\rangle$.

\subsubsection{Two-party expectation values}

We now describe the measurement of the two-party
expectation values $\langle A_1 B_0 \rangle$,
$\langle B_0 C_0 \rangle$, and $\langle B_0 C_1 \rangle$;
the corresponding circuits are shown in
Fig.~\ref{fig8}(b,c).
Because Alice's qubit $q_0$ and Bob's qubit $q_1$
originate from the same entangled source (and likewise
$q_2$ and Charlie's qubit $q_3$), these correlators do
not factorize into products of single-party expectation
values and must be measured jointly.

For $\langle A_1 B_0 \rangle$, qubits $q_0$, $q_1$, and
$q_2$ are measured directly in the computational basis
in a single circuit, with $q_0 \to c_2$, $q_1 \to c_1$,
and $q_2 \to c_0$, as shown in Fig.~\ref{fig8}(b).
The expectation value is the three-bit parity
\begin{equation}
    \langle A_1 B_0 \rangle = E[\{c_0, c_1, c_2\}]
    = \sum_{i,j,k} (-1)^{i+j+k}\, p_{ijk},
\end{equation}
where $p_{ijk}$ is the probability of the joint outcome
$|ijk\rangle$ of the three registers.

\begin{figure*}[t]
    \includegraphics[width=\linewidth]{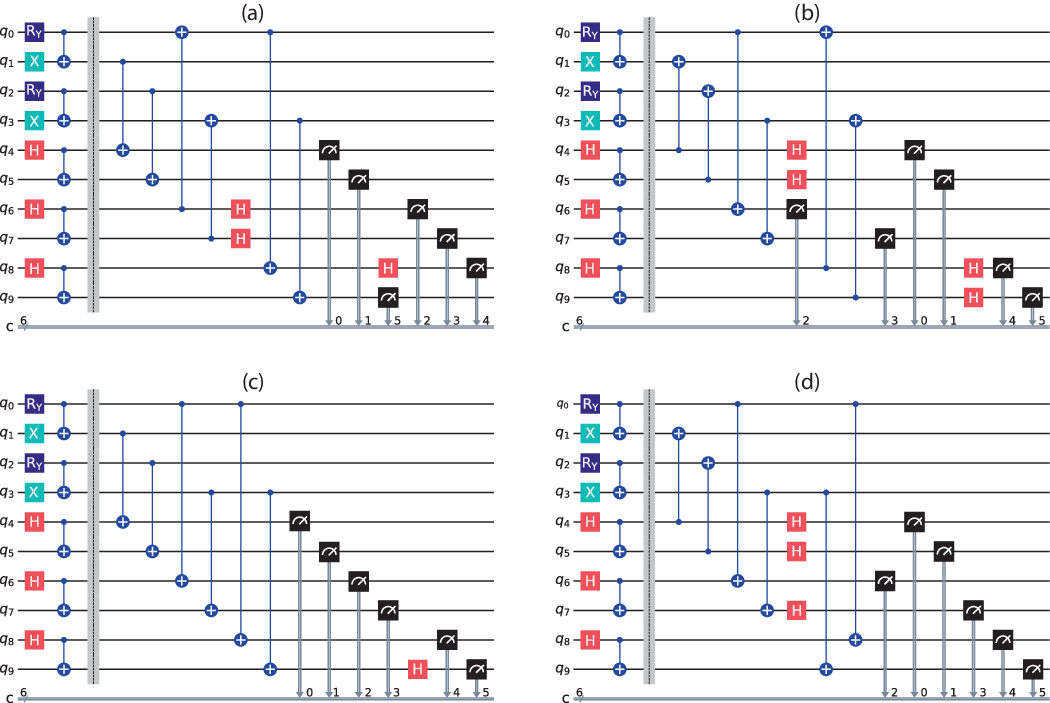}
    \caption{Circuits for measuring the three-party
    expectation values $\langle A_x B_y C_z \rangle$
    using six ancilla qubits $q_4$--$q_9$.
    (a) $\langle A_0 B_0 C_z \rangle$,
    (b) $\langle A_0 B_1 C_z \rangle$,
    (c) $\langle A_1 B_0 C_z \rangle$,
    (d) $\langle A_1 B_1 C_z \rangle$,
    for $z \in \{0,1\}$. The direction of CNOT
    gates and the placement of Hadamard gates are
    adjusted according to the target Pauli observable
    ($X$ or $Z$) of each party, following the
    convention of Sec.~\ref{appA2}. In all panels the
    meter qubits are measured into the classical
    registers in the fixed order $q_4 \to c_0$,
    $q_5 \to c_1$, $\ldots$, $q_9 \to c_5$.}
    \label{fig9}
\end{figure*}

For $\langle B_0 C_0 \rangle$ and $\langle B_0 C_1 \rangle$,
Bob's qubits $q_1$ and $q_2$ are measured in the
computational basis into registers $c_0$ and $c_1$, while
Charlie's qubit is read out simultaneously in the $Z$ and
$X$ bases with the ancilla-assisted scheme of
Fig.~\ref{fig7}(c), recording $q_3$ into $c_2$ and the
ancilla $q_4$ into $c_3$, as shown in
Fig.~\ref{fig8}(c). The two basis components,
\begin{align}
    \langle B_0 Z_C \rangle &= E[\{c_0, c_1, c_2\}], \\
    \langle B_0 X_C \rangle &= E[\{c_0, c_1, c_3\}],
\end{align}
are evaluated on the same shots, and both settings follow
as in Eq.~\eqref{eq:Cz_combination},
\begin{equation}
    \langle B_0 C_z \rangle = \frac{\langle B_0 Z_C \rangle
    + (-1)^z \langle B_0 X_C \rangle}{\sqrt{2}}, \quad z = 0, 1.
\end{equation}

\subsubsection{Three-party expectation values}

Finally, we describe the measurement of the three-party
expectation values $\langle A_x B_y C_z \rangle$ for
$x, y, z \in \{0,1\}$, which enter both the BLHV
correlators $\mathcal{I}$, $\mathcal{J}$ and the FNN witnesses
$\mathcal{R}_{\mathrm{CNS}}$, $\mathcal{R}_{\mathrm{NSC}}$. Each circuit
uses six meter qubits $q_4$--$q_9$, prepared as three
Bell-pair meters $(q_4, q_5)$, $(q_6, q_7)$, and
$(q_8, q_9)$ by the circuit of Fig.~\ref{fig6}(b), in
addition to the four system qubits $q_0$--$q_3$. The
four configurations are illustrated in Fig.~\ref{fig9}.
The pair $(q_4, q_5)$ reads Bob's joint observable
$B_y$, while the pairs $(q_6, q_7)$ and $(q_8, q_9)$
each read Alice together with Charlie, one with Charlie
coupled in the $Z$ basis and the other in the $X$ basis,
so that both settings $C_0$ and $C_1$ are obtained from
the same run. The direction of the CNOT couplings and
the placement of Hadamard gates select the target Pauli
observable of each party ($Z$: CNOT from system qubit to
meter; $X$: CNOT from meter to system qubit followed by
a Hadamard on the meter), following the convention
established in Sec.~\ref{appA2}.

In all four circuits, the meters are measured into the
classical registers in the fixed order $q_4 \to c_0$,
$q_5 \to c_1$, $\ldots$, $q_9 \to c_5$. A Bell-pair
meter carries information only in the joint parity of
its two halves; each correlator is therefore extracted
as a marginal parity of Bob's registers $\{c_0, c_1\}$
together with \emph{both} registers of one Alice-Charlie
meter pair,
\begin{align}
    E_{67} &= E[\{c_0, c_1, c_2, c_3\}], \label{eq:E67}\\
    E_{89} &= E[\{c_0, c_1, c_4, c_5\}]. \label{eq:E89}
\end{align}
Depending on the couplings of each configuration, one of
these parities is the $Z$-basis component
$\langle A_x B_y Z_C \rangle$ and the other the
$X$-basis component $\langle A_x B_y X_C \rangle$:
\begin{center}
\begin{tabular}{lcc}
\hline\hline
Configuration & $\langle A_x B_y Z_C \rangle$ & $\langle A_x B_y X_C \rangle$ \\
\hline
(a) $\langle A_0 B_0 C_z \rangle$ & $E_{89}$ & $E_{67}$ \\
(b) $\langle A_0 B_1 C_z \rangle$ & $E_{67}$ & $E_{89}$ \\
(c) $\langle A_1 B_0 C_z \rangle$ & $E_{67}$ & $E_{89}$ \\
(d) $\langle A_1 B_1 C_z \rangle$ & $E_{89}$ & $E_{67}$ \\
\hline\hline
\end{tabular}
\end{center}
Both settings of Charlie then follow, as in the single-
and two-party cases, from
\begin{equation}
    \langle A_x B_y C_z \rangle =
    \frac{\langle A_x B_y Z_C \rangle
    + (-1)^z \langle A_x B_y X_C \rangle}{\sqrt{2}},
    \label{eq:ABC_combination}
\end{equation}
for $z \in \{0,1\}$.

A fifth configuration measures $\langle A_0 \,(B_0 B_1)\, C_z \rangle$
with $B_0 B_1 = -Y \otimes Y$, which enters the reconstruction of
$\bar{B}_1$ in Eq.~\eqref{eq:B1bar}. It follows the same scheme, with
Bob's meter pair read in the $Y$ basis (an $S^\dagger$ gate followed by a
Hadamard before readout), the pair $(q_6, q_7)$ coupled as in
configuration (a), an overall sign $-1$, and the parities combined as in
Eq.~\eqref{eq:ABC_combination}.

We note that for the state family of Eq.~\eqref{eq:state}
the single-source cross-basis correlators vanish,
$\langle X \otimes Z \rangle_{\psi(\theta)} =
\langle Z \otimes X \rangle_{\psi(\theta)} = 0$ for all
$\theta$. Consequently, the component of
Eq.~\eqref{eq:ABC_combination} in which Alice's and
Charlie's bases differ, namely
$\langle A_0 B_y Z_C \rangle$ and
$\langle A_1 B_y X_C \rangle$, vanishes identically, and
the corresponding meter parity converges to zero
accordingly; the nonvanishing content of each correlator
is carried by the basis-matched component.

\bibliographystyle{apsrev4-2}
\bibliography{ref} 

\end{document}